\documentclass[preprint,aps,showkeys,floatfix]{revtex4}
\newcommand{\bfm}[1]{\mbox{\boldmath${#1}$}}
\begin{document}
\title{Canonical quantization of classical systems\\ with generalized entropies}
\author{A.M. Scarfone}\email{antonio.scarfone@polito.it}
\affiliation{Istituto Nazionale di Fisica della Materia (INFM) and Dipartimento di Fisica,\\
Politecnico di Torino, Corso Duca degli Abruzzi 24, I-10129
Torino, Italy}

\begin{abstract}
We present, in the framework of the canonical quantization, a
class of nonlinear Schr\"odinger equations with a complex
nonlinearity describing, in the mean field approximation, systems
of collectively interacting particles. The quantum evolution
equation is obtained starting from the study of a $N$-body
classical system where the underlined nonlinear kinetics is
governed by a kinetic interaction principle (KIP) recently
proposed [G. Kaniadakis: Physica A {\bf 296} (2001), 405--425].
The KIP, both imposes the form of the generalized entropy
associated to the classical system, and determines the
Fokker-Planck equation describing the kinetic evolution of the
system towards equilibrium.
\end{abstract}
 \maketitle

The Schr\"odinger equation is one of the most studied equations
both from a mathematical and a physical point of view. A
particular interest concerns the possible nonlinear extension of
this equation. Just one year after its discovery, Fermi proposed
the first nonlinear generalization \cite{Fermi}.

Among the many nonlinear extensions of the Schr\"odinger equation
(NSE) it is worthy to recall the effort made by Bialynicki-Birula
and Mycielski (BBM) \cite{Bialynicki} who proposed to introduce
the simple nonlinearity $-b\,\ln(|\psi|^2)\,\psi$, selected by
assuming the factorization of the wave function for composed
systems. Motivated by this work, some experimental tests were
carried out based on neutrons interferometry \cite{Shull}, and
Fresnel diffraction with slow neutrons \cite{Gaehler}. They gave
the upper limits $b<3.4\times10^{-13}$ and $b<3.3\times10^{-15}$,
respectively, suggesting that there is no a real ground for a such
nonlinear term in the Schr\"odinger equation.\\
Weinberg \cite{Weinberg}, in order to save partially the
superposition principle of the standard quantum mechanics,
proposed a very general class of NSEs with a homogeneous
nonlinearity. A further experimental test \cite{Bollinger} shows
again a sharper bound on the deviation from linearity.

Notwithstanding, many complex systems can be described by
introducing nonlinear terms in the linear Schr\"odinger equation,
to take into account the interactions among many particles.
Applications of the BBM equation to many particles system, in
particular in the nuclear physics, have been considered in
\cite{Hefter} (and references therein). There, it has been
suggested that, in spite of the negative tests based on neutrons,
the BBM equation and more generally, NSEs, can be usefully applied
in the description of extended objects.\\ The NSE with the cubic
nonlinearity \cite{Gross,Pitaevskii} has been used to study the
dynamical evolution of a boson gas with a $\delta$-function
pair-wise repulsion or attraction \cite{Barashenkov}, and in the
description of the Bose-Einstein condensation of alcali atoms like
$^7$Li, $^{23}$Na and $^{87}$Rb \cite{Stringari}.  In
\cite{Scarfone0} a NSE with complex nonlinearity has been
introduced in order to account for a generalized Pauli
exclusion-inclusion principle between the quantum particles of the
system. Finally it is worthy to mention the family of the
Doebner-Goldin equations \cite{Doebner}, introduced from
topological considerations, as the most general class of
Schr\"odinger equations compatible with the Fokker-Planck
equation, for the probability density $\rho=|\psi|^2$.

The BBM equation is closely connected with the Boltzmann-Gibbs
(BG) entropy. In fact, it can be shown that, within this theory,
the binding energy released in the splitting of an arbitrary wave
function, for an isolated system, in non-overlapping parts of the
same form $\psi({\bfm x})\to\sum_ip_{_i}^{1/2}\,\psi({\bfm
x}-{\bfm x}_{_i})$, is given by $\Delta\,E=-b\,\sum_ip_{_i}\,\ln
p_{_i}$, where $p_{_i}$ are normalized according to $\sum_i
p_{_i}=1$. In Ref. \cite{Hefter} it has been suggested that the
BBM equation can be used to formulate the thermodynamical
description of spatially extended or self-interacting quantum
mechanical objects.

In the present paper we approach to the canonical quantization of
a classical system, where the statistical information is supplied
by a very general entropy.

We recall that, in the recent years nonlinear generalizations of
the Fokker-Planck equation and their connection with generalized
entropies have attracted the interest of many authors
\cite{Compte,Frank1,Frank2}. Pure-electron plasma
\cite{Boghosian}, bremsstrahlung \cite{Tsallis2}, self-gravitating
systems \cite{PP} are few examples, among many others, where we
can find experimental evidence for the relevance of the use of
generalized entropies and nonlinear Fokker-Planck equations
(NFPEs).\\
In Ref. \cite{Kaniadakis1}, it has been proposed a kinetic
interaction principle (KIP) which defines a special collective
interaction among the $N$-identical particles of a classical
system. The KIP imposes the form of the generalized entropy and
governs the evolution of the system toward the equilibrium, by
fixing the expression of the nonlinear current appearing in the
NFPE.

We present the Lagrangian formulation of the model. By using the
minimum action principle, we obtain a class of nonlinear
Schr\"odinger equations containing a complex nonlinearity and
describing, in the mean field approximation, a quantum system of
interacting particles. The form of the entropy of the ancestor
classical system determines the nonlinearity in the evolution
equation. Finally, by means of a gauge transformation of third
kind \cite{Doebner,Scarfone1,Scarfone2} we show that this family
of NSEs is gauge equivalent to another family of
NSEs having a pure real nonlinearity depending only on the probability density field.\\
Some evolution equations obtained from kinetic equations known in
literature within the Boltzmann-Gibbs entropy, the Tsallis
entropy, the quantum-group entropy and the Kaniadakis entropy are
given to
illustrate the relevance of the method.\\

Let us consider a classical Markovian process in a $n$-dimensional
space
\begin{equation}
\frac{\partial\,\rho(t,{\bfm
x})}{\partial\,t}=\int\Big[\pi(t,{\bfm y}\rightarrow{\bfm
x})-\pi(t,{\bfm x}\rightarrow{\bfm y})\Big]\,d{\bfm y} \
,\label{ee}
\end{equation}
describing the kinetics of a system of $N$ identical particles. We
assume for the transition probability $\pi(t,{\bfm
x}\rightarrow{\bfm y})$ a suitable expression in terms of the
populations of the initial site $\bfm x$ and the final site $\bfm
y$.

According to KIP \cite{Kaniadakis1} we pose
\begin{equation}
\pi(t,{\bfm x}\rightarrow{\bfm y})=r(t,\,{\bfm x},\,{\bfm x}-{\bfm
y}) \,a(\rho)\,b(\rho^\prime)\,c(\rho,\,\rho^\prime) \
,\label{tran}
\end{equation}
where $\rho=\rho(t,\,{\bfm x})$ and
$\rho^\prime\equiv\rho(t,\,{\bfm y})$ are the particle density
functions in the starting site $\bfm x$ and in the arrival site
$\bfm y$, respectively, whereas $r(t,\,{\bfm x},\,{\bfm x}-{\bfm
y})$ is the transition rate which depends only on the starting
$\bfm x$ and arrival $\bfm y$ sites during the particle transition
${\bfm x}\rightarrow{\bfm y}$.\\ The two factors $a(x)$ and $b(x)$
in Eq. (\ref{tran}) are arbitrary non-negative functions, supposed
to be continuous. Their limit values for $x\to0$ are subject to
precise physical requirements. For instance, $a(0)=0$ because if
the starting site is empty the transition probability is equal to
zero; $b(0)=1$ because no inhibition occurs if the arrival site is
empty. Finally, the last factor $c(\rho,\,\rho^\prime)$ takes into
account that the populations of the two sites can eventually
affect the transition, collectively and symmetrically.

The expression of the functional $b(\rho)$ plays a very important
role in the particle kinetics because can stimulate or inhibit the
transition ${\bfm x}\rightarrow{\bfm y}$ allowing in this way to
take into account interactions originated from collective
effects.

With the assumption given in Eq. (\ref{tran}) for the transition
probability, under the first neighbor approximation, we can expand
the r.h.s of Eq. (\ref{ee}) up to the second order in $\bfm y$
\cite{Kaniadakis1}
\begin{equation}
\frac{\partial\,\rho}{\partial\,t}=\frac{\partial}{\partial
x_{_i}}\left[\left(
\zeta_{_i}+\frac{\partial\,\zeta_{_{ij}}}{\partial
x_{_j}}\right)\,\gamma(\rho)+\zeta_{_{ij}}\,\gamma(\rho)\,
\frac{\partial}{\partial\,x_{_j}}\,\ln\kappa(\rho)\right] \
,\label{kip}
\end{equation}
with $i=1,\,\cdots,\,n$ and summation over repeated indices is
assumed.\\ In Eq. (\ref{kip}) $\gamma(\rho)\equiv
a(\rho)\,b(\rho)\,c(\rho,\,\rho)$ and
$\kappa(\rho)=a(\rho)/b(\rho)$, whereas the coefficients
$\zeta_{_i}$ and $\zeta_{_{ij}}$ are given by
\begin{equation}
\zeta_{_i}=\int r(t,\,{\bfm x},\,{\bfm y})\,y_{_i}\,d{\bfm y} \
,\hspace{10mm}{\rm and}\hspace{10mm} \zeta_{_{ij}}=\frac{1}{2}\int
r(t,\,{\bfm x},\,{\bfm y})\,y_{_i}\,y_{_j}\,d{\bfm y} \
.\label{ees}
\end{equation}

In the following we consider the case of linear drift by posing
$\gamma(\rho)=\rho$ and after introducing the quantity ${\bfm
u}_{\rm
drift}|_{_i}=-\zeta_{_i}-\partial\,\zeta_{_{ij}}/\partial\,x_{_j}$,
the $i$-th component of the drift velocity and assuming
$\zeta_{_{ij}}=D\,\delta_{_{ij}}$, with $D$ the diffusion
coefficient, Eq. (\ref{kip}) becomes
\begin{equation}
\frac{\partial\,\rho}{\partial\,t}+{\bfm
\nabla}\cdot\Big\{\rho\,\Big[{\bfm u}_{\rm
drift}-D\,{\bfm\nabla}\,\ln\kappa(\rho)\Big]\Big\}=0 \
,\label{cont2}
\end{equation}
which is a NFPE for the field $\rho$. The total current ${\bfm
j}={\bfm j}_{\rm drift}+{\bfm j}_{\rm diff}$ is the sum of two
quantities: ${\bfm j}_{\rm drift}=\rho\,{\bfm u}_{\rm drift}$
which is the standard linear drift current and ${\bfm j}_{\rm
diff}=-D\,\rho\,{\bfm \nabla}\ln\kappa(\rho)$ which is a nonlinear
diffusion current different from the standard Fick's one.

Eq. (\ref{cont2}) describes a class of nonlinear diffusive
processes varying the functional $\kappa(\rho)$. In the following
we assume a constant diffusion coefficient $D=const.$

The functional $\kappa(\rho)$ is related to the entropy of the
classical system as can be showed in the non-equilibrium
thermodynamics framework \cite{Prigogine,deGroot}, by relating the
production of the entropy to a NFPE. In fact, we may show this
link on a more general ground by starting from a constrained
entropic form
\begin{equation}
{\mathcal S(\rho)}=S(\rho)-\beta\int {\mathcal E}({\bfm
x})\,\rho\,d{\bfm x}-\beta^\prime\int \rho\,d{\bfm x} \
,\label{funentropy}
\end{equation}
where $\beta$ and $\beta^\prime$ are the Lagrange multipliers
associated to the constraints of the total energy $E=\int{\mathcal
E}({\bfm x})\,\rho\,d{\bfm x}$ and the normalization $\int
\rho\,d{\bfm x}=1$.

Without loss of generality, we can assume for the entropy function
the following expression
\begin{equation}
S(\rho)=-k\int d{\bfm x}\int d\rho\,\ln\kappa(\rho) \
.\label{entropy}
\end{equation}
The standard BG-entropy is recovered from Eq. (\ref{entropy})
after posing $\kappa(\rho)=e\,\rho$.

Quite generally, the evolution of the field $\rho$ in the
configuration space is governed by the continuity equation
\begin{equation}
\frac{\partial\,\rho}{\partial\,t}+{\bfm \nabla}\cdot{\bfm J}=0 \
,\label{cont}
\end{equation}
and we assume a linear relation between the current $\bfm J$ and
the constrained thermodynamic force ${\bfm\nabla}(\delta{\mathcal
S}/\delta\rho)$
\begin{equation}
{\bfm
J}=\frac{D}{k}\,\rho\,{\bfm\nabla}\left(\frac{\delta{\mathcal
S}}{\delta\rho}\right) \ .\label{d}
\end{equation}
Putting Eq. (\ref{d}) into Eq. (\ref{cont}) we obtain
\begin{equation}
\frac{\partial\,\rho}{\partial\,t}+{\bfm
\nabla}\cdot\left\{\rho\left[-\frac{D}{k}\,\beta\,{\bfm
\nabla}\,{\mathcal E}({\bfm
x})-D\,{\bfm\nabla}\,\ln\kappa(\rho)\right]\right\}=0 \
,\label{cont1}
\end{equation}
and, after introducing the drift velocity
\begin{equation}
{\bfm u}_{\rm drift}=-\frac{D}{k}\,\beta\,{\bfm\nabla}\,{\mathcal
E}({\bfm x}) \ ,
\end{equation}
Eq. (\ref{cont1}) takes the same form of the continuity equation
(\ref{cont2}).

Let us now introduce the wave function $\psi$ describing, in the
mean field approximation, the quantum analogous of the classical
$N$-particle interacting system depicted above. We postulate that
the evolution equation of this system is given by a NSE
\begin{eqnarray}
i\,\hbar\,\frac{\partial\,\psi}{\partial\,t}=
-\frac{\hbar^2}{2\,m}\,\Delta\,\psi+\Lambda(\psi^\ast,\,\psi)\,\psi+V({\bfm
x})\,\psi \ ,\label{schroedinger1}
\end{eqnarray}
where $\Lambda(\psi^\ast,\,\psi)$ is a complex nonlinearity and
$V({\bfm x})$ is the external potential.

In the canonical picture, the motion equation
(\ref{schroedinger1}) can be obtained after introducing the action
of the system
\begin{equation}
{\mathcal A}=\int{\mathcal L}(\psi^\ast,\,\psi)\,d{\bfm x}\,dt \
,\label{az}
\end{equation}
where ${\mathcal L}$ is the Lagrangian density given by
\begin{equation}
{\mathcal
L}(\psi^\ast,\,\psi)=i\,\frac{\hbar}{2}\,\left(\psi^\ast\,\frac{\partial\,\psi}{\partial\,t}
-\frac{\partial\,\psi^\ast}{\partial\,t}\,\psi\right)-
\frac{\hbar^2}{2\,m}\,|
{\bfm\nabla}\,\psi|^2-U(\psi^\ast,\,\psi)-\psi^\ast\,V({\bfm
x})\,\psi \ ,\label{lag1}
\end{equation}
and $U(\psi^\ast,\,\psi)$ is a nonlinear potential describing the
collective interaction between the particles.

The NSE (\ref{schroedinger1}) is obtained by applying the minimum
action principle to the system described by the Lagrangian
(\ref{lag1})
\begin{equation}
\frac{\delta\,{\mathcal A}}{\delta\,\psi^\ast}=0 \ .\label{psi}
\end{equation}
The nonlinear functional $\Lambda(\psi^\ast,\,\psi)$ in Eq.
(\ref{schroedinger1}) is given by
\begin{equation}
\Lambda(\psi^\ast,\,\psi)=\frac{\delta}{\delta\,\psi^\ast}\int
U(\psi^\ast,\,\psi)\,d{\bfm x}\,dt \ .\label{lambda}
\end{equation}

It is convenient at this point to introduce the hydrodynamic
fields $\rho$ and $\Sigma$ related to the $\psi$-function through
the polar decomposition \cite{Bohm,Madelung}
\begin{eqnarray}
\psi(t,\,{\bfm x})=\rho^{1/2}(t,\,{\bfm
x})\,\exp\left(\frac{i}{\hbar}\,\Sigma(t,\,{\bfm x})\right) \
.\label{polar}
\end{eqnarray}
Introducing the real part $W(\psi^\ast,\,\psi)$ and the imaginary
part ${\mathcal W}(\psi^\ast,\,\psi)$ of the complex nonlinearity
$\Lambda(\psi^\ast,\,\psi)=W(\psi^\ast,\,\psi)+i\,{\mathcal
W}(\psi^\ast,\,\psi)$, Eq. (\ref{schroedinger1}) can be separated
in a coupled of real nonlinear evolution equations for the phase
$\Sigma$
\begin{equation}
\frac{\partial\,\Sigma}{\partial\,t}+\frac{\left({\bfm\nabla}\,\Sigma\right)^2}{2\,m}
-\frac{\hbar^2}{2\,m}\,\frac{\Delta\sqrt{\rho}}{\sqrt{\rho}}+W(\rho,\,\Sigma)+V({\bfm
x})=0 \ ,\label{s}
\end{equation}
and the density $\rho$
\begin{equation}
\frac{\partial\,\rho}{\partial\,t}+{\bfm\nabla}\cdot\left(\frac{{\bfm
\nabla}\Sigma}{m}\,\rho\right)-\frac{2}{\hbar}\,\rho\,{\mathcal
W}(\rho,\,\Sigma)=0 \ .\label{rho}
\end{equation}
In Eq. (\ref{rho}) ${\bfm
j}_{_0}\equiv({\bfm\nabla}\,\Sigma/m)\,\rho$ is the linear current
of the standard quantum mechanics.

Both Eq.s (\ref{s}) and (\ref{rho}) can be obtained from the
variational problem
\begin{equation}
\frac{\delta\,{\mathcal A}}{\delta\,\rho}=0 \ ,\hspace{10mm}{\rm
and}\hspace{10mm} \frac{\delta\,{\mathcal A}}{\delta\,\Sigma}=0 \
,\label{rhos}
\end{equation}
where the Lagrangian density (\ref{lag1}), after taking into
account Eq. (\ref{polar}), assumes the expression
\begin{equation}
{\mathcal
L}(\rho,\,\Sigma)=-\frac{\partial\,\Sigma}{\partial\,t}\,\rho-
\frac{({\bfm\nabla}\,\Sigma)^2}{2\,m}\,\rho-\frac{\hbar^2}{8\,m}
\,\frac{\left({\bfm\nabla}\,\rho\right)^2}{\rho}-U(\rho,\,\Sigma)-V({\bfm
x})\,\rho \ .\label{ham1}
\end{equation}
We observe that the nonlinear functionals $W(\rho,\,\Sigma)$ and
${\mathcal W}(\rho,\,\Sigma)$ are related to the nonlinear
potential $U(\rho,\,\Sigma)$ through the relations
\begin{equation}
W(\rho,\,\Sigma)=\frac{\delta}{\delta\,\rho}\int
U(\rho,\,\Sigma)\,d{\bfm x}\,dt \ ,\hspace{10mm} {\mathcal
W}(\rho,\,\Sigma)=\frac{\hbar}{2\,\rho}\,\frac{\delta}{\delta\,\Sigma}\int
U(\rho,\,\Sigma)\,d{\bfm x}\,dt \ ,\label{cw}
\end{equation}
and by choosing $U(\rho,\,\Sigma)$ as
\begin{equation}
U(\rho,\,\Sigma)=-D\,\rho\,{\bfm\nabla}\,\ln
\kappa(\rho)\cdot{\bfm\nabla \Sigma} \ ,\label{u}
\end{equation}
from Eq.s (\ref{s}) and (\ref{rho}) we obtain
\begin{eqnarray}
&&\frac{\partial\,\Sigma}{\partial\,t}+
\frac{({\bfm\nabla\,\Sigma})^2}{2\,m}+\frac{\hbar^2}{2\,m}\,
\frac{\Delta\,\sqrt{\rho}}{\sqrt{\rho}}+D\,\rho\,\frac{\partial}{\partial\,\rho}\ln
\kappa(\rho)\,\Delta \Sigma+V({\bfm x})=0
\ ,\label{s1}\\
&&\frac{\partial\,\rho}{\partial\,t}+{\bfm\nabla}\cdot
\left[\frac{{\bfm\nabla}\,\Sigma}{m}\,\rho-D\,\rho\,{\bfm\nabla}\ln\kappa(\rho)\right]=0
\ .\label{r1}
\end{eqnarray}
By posing $\widehat{\bfm u}_{\rm drift}={\bfm\nabla}\,\Sigma/m$,
the quantum drift velocity, Eq. (\ref{r1}) assumes the same
expression of the Fokker-Planck equation (\ref{cont2}) describing
the kinetics of the classical system obeying to KIP.

The NSE in the $\psi$-representation is given by
\begin{eqnarray}
\nonumber &&i\,\hbar\,\frac{\partial\,\psi}{\partial\,t}=
-\frac{\hbar^2}{2\,m}\,\Delta\,\psi +
D\,\rho\,\frac{\partial}{\partial\,\rho} \ln
\kappa(\rho)\,\Delta\,\Sigma\,\psi+i\,\frac{\hbar\,D}{2\,\rho}\,{\bfm\nabla}\Big[\rho
\,{\bfm\nabla}\ln\kappa(\rho)\Big]\,\psi+V({\bfm x})\,\psi \ ,\\
\label{schroedinger2}
\end{eqnarray}
where the nonlinearities $W$ and $\mathcal W$ are determined
through the functional $\kappa(\rho)$ which also defines the
entropy (\ref{entropy}) of the ancestor classical system.

We introduce now the following nonlinear gauge transformation of
third kind \cite{Scarfone1,Scarfone2}
\begin{equation}
\psi\rightarrow\phi=\psi\,\exp\left(-\frac{i}{\hbar}\,m\,D\,\ln\kappa(\rho)\right)
\ ,\label{gauge}
\end{equation}
which, being an unitary transformation, does not change the
amplitude of the wave function, $|\psi|^2=|\phi|^2=\rho$, and
transforms the phase $\Sigma$ of the old field $\psi$ in the
phases $\sigma$ of the new field $\phi$ according to the equation
\begin{equation}
\sigma=\Sigma-m\,D\,\ln\kappa(\rho) \ .\label{phase}
\end{equation}

After performing the transformation (\ref{gauge}), Eq.
(\ref{schroedinger2}) becomes
\begin{eqnarray}
\nonumber
&&\hspace{-10mm}i\,\hbar\,\frac{\partial\,\phi}{\partial\,t}=
-\frac{\hbar^2}{2\,m}\,\Delta\,\phi+
m\,D^2\left\{\rho\,\frac{\partial}{\partial\,\rho} \ln
\kappa(\rho)\,\Delta\,\ln \kappa(\rho)+{1\over2}\,
\Big[{\bfm\nabla}\ln\kappa(\rho)\Big]^2\right\}\,\phi+V({\bfm
x})\,\phi \ ,\\ \label{schroedinger3}
\end{eqnarray}
where in Eq. (\ref{schroedinger3}) appears a purely
real nonlinearity depending only on the field $\rho$.\\

To illustrate the relevance of the approach described above, we
derive some NSEs starting from as much entropies known in
literature: BG-entropy, Tsallis-entropy, quantum-group entropy and
$\kappa$-entropy.

(a) As a first example we consider the standard classical entropy
\begin{equation}
S_{{\rm BG}}(\rho)=-k\int\rho\,\ln\rho\,d{\bfm x} \ ,\label{SBG}
\end{equation}
which follows from Eq. (\ref{entropy}) by posing
$a(\rho)=e\,\rho$, $b(\rho)=1$ so that $\kappa(\rho)=e\,\rho$. The
corresponding continuity equation becomes
\begin{equation}
\frac{\partial \rho}{\partial
t}+{\bfm\nabla}\cdot\left(\frac{{\bfm
\nabla}\,\Sigma}{m}\,\rho-D\, {\bfm\nabla}\,\rho\right)=0 \
,\label{Shannon}
\end{equation}
that is the linear Fokker-Planck equation \cite{Doebner}.

The NSE is readily obtained from Eq. (\ref{schroedinger3}) and
becomes
\begin{equation}
i\,\hbar\,\frac{\partial\,\phi}{\partial\,t}=-\frac{\hbar^2}{2\,m}\,\Delta\,\phi
+m\,D^2\,\left[\frac{\Delta\,\rho}{\rho}
-{1\over2}\left(\frac{{\bfm\nabla}\,\rho}{\rho}\right)^2\right]\,\phi+V({\bfm
x})\,\phi \ ,\label{DG1}
\end{equation}
which was studied previously in \cite{Guerra}. In particular, Eq.
(\ref{DG1}) is equivalent to the linear Schr\"odinger equation
\begin{equation}
i\,k^{\!\!\!\!\!-}\frac{\partial\,\chi}{\partial\,t}=-\frac{{k^{\!\!\!\!\!-}}^2}
{2\,m}\,\Delta\,\chi+V({\bfm x})\,\chi \ ,
\end{equation}
with $k^{\!\!\!\!\!-}=\hbar\,\sqrt{1-(2\,m\,D/\hbar)^2}$ and the
field $\chi$ is related to the hydrodynamic fields $\rho$ and
$\sigma$ through the relation $\chi=\rho^{1/2}\,
\exp(i\,\sigma/k^{\!\!\!\!\!-})$.

(b) As a second example we consider the Tsallis entropy
\cite{Tsallis}
\begin{equation}
S_{_q}(\rho)=k\,\int\frac{\rho^q-\rho}{1-q}\,d{\bfm x} \
.\label{tsallis}
\end{equation}
Eq. (\ref{tsallis}) can be obtained from Eq. (\ref{entropy}) after
posing
\begin{equation}
\ln\kappa(\rho)=-\ln_{_q}\left(\frac{\alpha}{\rho}\right) \ ,
\end{equation}
where $\alpha=q^{1/(1-q)}$ is a constant and the deformed
$q$-logarithm is defined by
\begin{equation}
\ln_{_q}(x)=\frac{x^{1-q}-1}{1-q} \ ,
\end{equation}
which reduces to the standard logarithm for $q\to1$
($\ln_{_1}(x)=\ln x$) as well as Eq. (\ref{tsallis}) reduces to
the BG-entropy in the same limit.

The evolution equation for the field $\rho$ is given by
\begin{equation}
\frac{\partial\,\rho}{\partial\,t}+{\bfm\nabla}\cdot\left(\frac{{\bfm
\nabla}\,\Sigma}{m}\,\rho-D\,{\bfm\nabla}\,\rho^q\,\right)=0 \
,\label{Tsallis1}
\end{equation}
which is the NFPE discussed in Ref. \cite{Compte1,Frank1}.

The NSE associated to Eq. (\ref{Tsallis1}) assumes the form
\begin{eqnarray}
i\,\hbar\,\frac{\partial\,\phi}{\partial\,t}=-\frac{\hbar^2}{2\,m}\,
\Delta\,\phi+q^2\,m\,D^2\,\rho^{2\,(q-1)}\,\left[\frac{\Delta\,\rho}{\rho}
-\left({3\over2}-q\right)\,\left(\frac{{\bfm\nabla}\,\rho}{\rho}\right)^2\right]\,\phi+V({\bfm
x})\,\phi \ , \label{qDGd}
\end{eqnarray}
and reduces to Eq. (\ref{DG1}) for $q\to1$.

We observe that the nonlinearity in Eq. (\ref{qDGd}) coincides
with the same one appearing in the NSE proposed in Ref.
\cite{Olavo}, after replacing $q\rightarrow2-q$, and obtained in
the Tsallis statistics framework by using a different approach.

(c) Another interesting example is provided by the quantum groups
entropy \cite{Abe}
\begin{equation}
S_{_q}(\rho)=-k\,\int\frac{\rho^q-
\rho^{q^{-1}}}{q-q^{-1}}\,d{\bfm x} \ ,\label{sa}
\end{equation}
obtained from Eq. (\ref{entropy}) with the choice
\begin{equation}
\ln\kappa(\rho)=\frac{q\,\rho^{q-1}-q^{-1}\,\rho^{q^{-1}-1}}{q-q^{-1}}
\ .
\end{equation}
Entropy (\ref{sa}) is invariant under the interchange
$q\leftrightarrow q^{-1}$ and reduces to the BG-entropy in the
$q\to1$ limit. \\
The evolution equation for the field $\rho$ is given by
\begin{equation}
\frac{\partial\,\rho}{\partial\,t}+{\bfm\nabla}\cdot
\left[\frac{{\bfm
\nabla}\,\Sigma}{m}\,\rho-D\,{\bfm\nabla}\,\left(\frac{q}{q+1}\,
\rho^q+\frac{q^{-1}}{q^{-1}+1}\,\rho^{q^{-1}}\right)\,\right]=0 \
,\label{abe}
\end{equation}
and the associated NSE is readily obtained from Eq.
(\ref{schroedinger3}) and assumes the form
\begin{eqnarray}
\nonumber
i\,\hbar\,\frac{\partial\,\phi}{\partial\,t}=-\frac{\hbar^2}{2\,m}\,
\Delta\,\phi+\frac{m\,D^2}{(q-q^{-1})^2}\,\Bigg\{\rho\,
\frac{\partial\,f_{_q}(\rho)}{\partial\,\rho}\,\Delta\,f_{_q}(\rho)+{1\over2}\,
\Big[{\bfm\nabla}\,f_{_q}(\rho)\Big]^2\Bigg\}\,\phi+V({\bfm
x})\,\phi \ ,\\ \label{aDGd}
\end{eqnarray}
where $f_{_q}(\rho)=q\,\rho^{q-1}-q^{-1}\,\rho^{q^{-1}-1}$.

(d) As a final example let us consider the $\kappa$-entropy
\cite{Kaniadakis1,Kaniadakis3}
\begin{equation}
S_{_{\{\kappa\}}}(\rho)=-k\,\int\frac{\rho^{1+\kappa}-
\rho^{1-\kappa}}{2\,\kappa}\,d{\bfm x} \ ,\label{sk}
\end{equation}
which can be obtained from Eq. (\ref{entropy}) after posing
\begin{equation}
\ln\kappa(\rho)=\lambda\,\ln_{_{\{{\scriptstyle
\kappa}\}}}\left(\frac{\rho}{\alpha}\right) \ .
\end{equation}
The two quantities $\lambda=\sqrt{1-\kappa^2}$ and
$\alpha=[(1-\kappa)/(1+\kappa)]^{1/2\,\kappa}$ are constants and
fulfil the relation $(1\pm\kappa)\,\alpha^{\pm\kappa}=\lambda$.

The $\kappa$-logarithm is defined as
\begin{equation}
\ln_{_{\{\kappa\}}}(x)=\frac{x^\kappa-x^{-\kappa}}{2\,\kappa} \
,\label{klog}
\end{equation}
and reduces, in the $\kappa\to0$ limit, to the standard logarithm
$(\ln_{_{\{0\}}}(x)=\ln x)$.

From Eq. (\ref{r1}) we obtain the following continuity equation
for the field $\rho$
\begin{equation}
\frac{\partial\,\rho}{\partial\,t}+{\bfm\nabla}\cdot
\left[\frac{{\bfm
\nabla}\,\Sigma}{m}\,\rho-{D\over2}\,{\bfm\nabla}\,\left(
\rho^{1+\kappa}+\rho^{1-\kappa}\right)\right]=0 \ ,\label{kappa}
\end{equation}
which coincides with that proposed in Ref. \cite{Kaniadakis1}.

The associated NSE is given by
\begin{eqnarray}
\nonumber
&&i\,\hbar\,\frac{\partial\,\phi}{\partial\,t}=-\frac{\hbar^2}{2\,m}\,
\Delta\,\phi+{\lambda\over2}\,m\,D^2\,\left\{\left[(1+\kappa)\,\rho^\kappa
+(1-\kappa)\,\rho^{-\kappa}\right]\,
\Delta\,\ln_{_{\{\kappa\}}}\left(\frac{\rho}{\alpha}\right)\right.\\
&&\hspace{15mm}+\left.\lambda\,\left[{\bfm\nabla}\,\ln_{_{\{\kappa\}}}\left(\frac{\rho}{\alpha}\right)
\right]^2\right\}\,\phi+V({\bfm x})\,\phi \ ,\label{kDGd}
\end{eqnarray}
and reduces to the Doebner-Golding Eq. (\ref{DG1}) in the
$\kappa\to0$ limit as well as Eq. (\ref{sk}) reduces to the standard BG-entropy.\\


\end{document}